\documentclass[twocolumn,english,pra,preprintnumbers,showpacs,nofootinbib]{revtex4}
\usepackage{graphicx}
\usepackage{babel}

\newcommand{\lp}{\left(}
\newcommand{\rp}{\right)}
\newcommand{\nn}{\nonumber}
\newcommand{\al}{\alpha}

\begin{document}

\title{\boldmath Radiative-nonrecoil corrections of order $\al^{2}(Z\al)^{5}$
to the Lamb shift}

\preprint{Alberta Thy 17-09}

\author{Matthew Dowling}
\author{Jorge Mond{\'e}jar}
\author{Jan H. Piclum}
\author{Andrzej Czarnecki}
\affiliation{Department of Physics, University of Alberta, Edmonton, Alberta,
Canada T6G 2G7}

\date{\today}
\begin{abstract}
We present results for the corrections of order $\al^{2}(Z\al)^{5}$
to the Lamb shift. We compute all the contributing Feynman diagrams
in dimensional regularization and a general covariant gauge using
a mixture of analytical and numerical methods. We confirm results
obtained by other groups and improve their precision. Values of the
32 {}``master integrals'' for this and similar problems are provided.
\end{abstract}


\maketitle

\section{Introduction}

Recent developments in spectroscopy have led to very precise experimental
values for the $1S$ Lamb shift and the Rydberg
constant~\cite{Berkeland:1995zz,Weitz:1995zz,Bourzeix:1996zz,Udem:1997zz,Schwob:1999zz},
so that now the Lamb shift provides the best test of Quantum Electrodynamics
for an atom. These achievements have spurred great theoretical efforts
aimed at matching the current experimental accuracy (for a review
of the present status and recent developments in the theory of light
hydrogenic atoms, see~\cite{Eides:2000xc}).

The theoretical prediction
is expressed in terms of three small parameters: $Z\alpha$ describing
effects due to the binding of an electron to a nucleus of atomic number
$Z$; $\alpha$ (frequently accompanied by $1/\pi$) from electron
selfinteractions; and the ratio of electron to nucleus masses. The
Lamb shift is of the order $\alpha\left(Z\alpha\right)^{4}$; all
corrections through the second order in the small parameters are known,
as well as some of the third order \cite{Eides:2006hg}. 

Another source of corrections is the spatial distribution of the nuclear
charge. Even for hydrogen, the experimental uncertainty in the measurement
of the proton root mean square charge radius poses an obstacle for
further theoretical progress. Fortunately, measurements can be performed
also with the muonic hydrogen whose spectrum is much more sensitive
to the proton radius. A comparison of the theoretical prediction \cite{Pachucki:1996}
and anticipated new measurements \cite{muonicH} is expected to soon
improve the knowledge of this crucial parameter. 

In this paper we focus on the second-order radiative-nonrecoil contributions
to the Lamb shift of order $\al^{2}(Z\al)^{5}$. The total result
for the corrections of this order was presented first in \cite{Pachucki:1994zz}
and improved in \cite{Eides:1995gy,Eides:1995ey}. Our full result
is compatible with the previous ones and has better precision. When
comparing contributions from individual diagrams with \cite{Eides:1995ey},
however, we find small discrepancies in some cases.

In Section~\ref{eval} we present the details of our approach, and
in Section~\ref{res} we present our results. In Appendix~\ref{app}
we show the results for the master integrals, and in Appendix~\ref{appB}
new analytic results for two diagrams.

\section{Evaluation}
\label{eval}

We consider an electron of mass $m$ orbiting a nucleus
of mass $M$ and atomic number $Z$, where $Z$ is assumed to be of
such a size that $Z\al$ is a reasonable expansion parameter. We are
interested in corrections to the Lamb shift of order $\al^{2}(Z\al)^{5}$
and leading order in $m/M$, given by
\begin{equation}
  \delta E=-|\psi_{n}(0)|^{2}\mathcal{M}^{(2,2,0)}(eN\to eN)\,,
\end{equation}
where $|\psi_{n}(0)|^{2}=(Z\al\mu)^{3}/(\pi n^{3})$ is the squared
modulus of the wave function of an $S$ bound state with principal
quantum number $n$ ($\mu$ is the reduced mass of the system), and
$\mathcal{M}^{(2,2,0)}(eN\to eN)$ is the momentum space representation of
the amplitude of the interaction between the electron and the nucleus
at orders $\al^{2}(Z\al)^{2}$ and $(m/M)^{0}$. Both particles are
considered to be at rest and on their mass shell~\cite{Peskin}.

\begin{figure}[t]
  \includegraphics[width=\columnwidth]{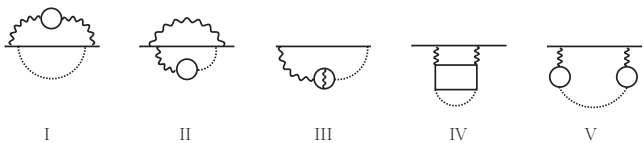}
  \caption{\label{vacuum}The different sets of vacuum polarization
      diagrams. Each set represents the drawn diagram plus all the
      possible permutations of its pieces.}
\end{figure}

\begin{figure}[t]
  \includegraphics[width=\columnwidth]{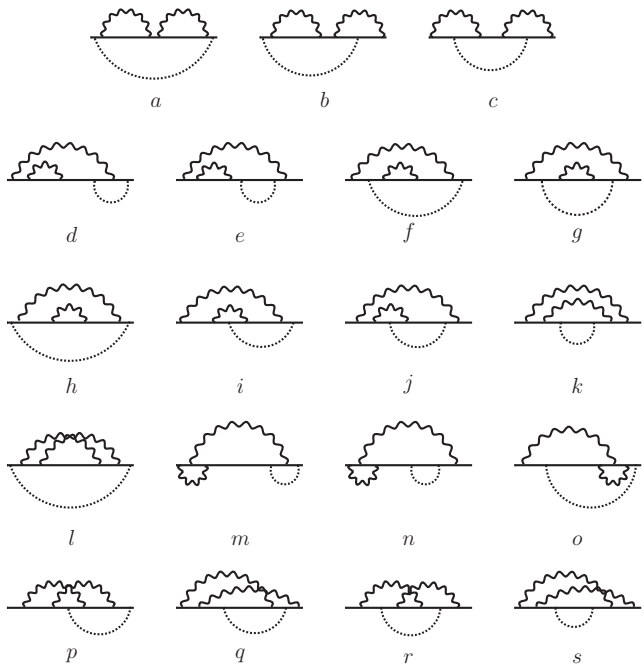}
  \caption{\label{other}The diagrams involving a two-loop electron
    self-interaction and vertex corrections.}
\end{figure}

\begin{figure}[t]
  \includegraphics[width=\columnwidth]{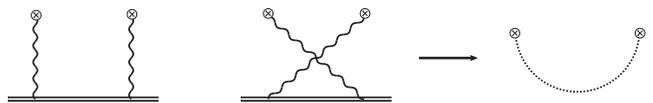}
  \caption{\label{graphdelta}The sum of the direct and crossed diagrams
    is approximated by an effective propagator (the double line
    represents the propagator of the nucleus).}
\end{figure}

The correction $\delta E$ is given by the sum of all the three-loop
diagrams presented in Figs.~\ref{vacuum} and \ref{other}. In these
figures, the continuous line represents the electron, and the dashed
line represents the interaction with the nucleus. The reason for this
is that for our purposes this interaction can be replaced by an effective
propagator. In all diagrams, the leading order in $m/M$ comes from
the region where all the loop momenta scale like $m$. The part of
the diagrams representing the interaction between the electron and
the nucleus at order $(Z\al)^{2}$ is given by the sum of the direct
and crossed two-photon exchange shown in Fig.~\ref{graphdelta}.
If $k$ and $N$ are the loop and nucleus momenta, respectively, and
$k^{2}\sim m^{2}\ll N^{2}=M^{2}$, the sum of the nucleus propagators
can be approximated at leading order by
\begin{eqnarray}
  &  & \frac{1}{(N+k)^{2}-M^{2}+i\epsilon} +
  \frac{1}{(N-k)^{2}-M^{2}+i\epsilon} \nn\label{delta}\\
  &  & \simeq\frac{1}{2N\cdot k+i\epsilon} - 
  \frac{1}{2N\cdot k-i\epsilon}=-i\pi\delta(N\cdot k)\,.
\end{eqnarray}
Since the nucleus is considered to be at rest, this gives us a $\delta(Mk^{0})$.
Together with the propagators of the two photons, this constitutes
the effective propagator.

We used dimensional regularization, and renormalized our results using
the on-shell renormalization scheme. For all the photon propagators
in Figs.~\ref{vacuum} and \ref{other} we used a general covariant
$R_{\xi}$ gauge. The overall cancellation of the dependence on the
gauge parameter in the final result provided us with a good check
for our calculations. Since the gauge invariance of the sum of the
subdiagrams in Fig.~\ref{graphdelta} is trivially and independently
fulfilled, for the photonic part of the effective propagator we used
the Feynman gauge.

Since we are considering free asymptotic states with independent spins,
the Dirac structures of the electron and the nucleus factorize, and
we simplify them by inserting each of the structures between the spinors
of the initial and final states and averaging over the spins of the
initial states.

We use the program \texttt{qgraf}~\cite{Nogueira:1991ex} to generate
all of the diagrams, and the packages \texttt{q2e} and
\texttt{exp}~\cite{Harlander:1997zb,Seidensticker:1999bb}
to express them as a series of vertices and propagators that can be
read by the \texttt{FORM}~\cite{Vermaseren:2000nd} package \texttt{MATAD~3}~\cite{Steinhauser:2000ry}.
Finally, \texttt{MATAD~3} is used to represent the diagrams in terms
of a set of scalar integrals using custom-made routines. In this way,
we come to represent the amplitude $\mathcal{M}$ in terms of about
18000 different scalar integrals. These integrals can be expressed
in terms of a few master integrals by means of integration-by-parts
(IBP) identities~\cite{Chetyrkin:1981qh}. Using the so-called Laporta
algorithm \cite{Laporta:1996mq,Laporta:2001dd} as implemented in
the \texttt{Mathematica} package \texttt{FIRE}~\cite{Smirnov:2008iw},
we find 32 master integrals.%
\footnote{Actually, it is possible to reduce the number of integrals to at least
31, and possibly 30. We give more details about this in Appendix \ref{app}.}

Since the program \texttt{FIRE} deals only with standard propagators,
when using it we worked only with one of the nucleus propagators,
instead of the Dirac delta of the effective propagator. That is, instead
of working with $\delta(k^{0})$, we worked with $1/(2k\cdot N+i\epsilon)$,
for example. Working with just one of the propagators is enough for
this purpose, as the IBP method is insensitive to the $i\epsilon$
prescription (remember from Eq.~(\ref{delta}) that in our approximation
this is the only difference between the two propagators). Since each
diagram in Figs.~\ref{vacuum} and \ref{other} represents the subtraction
of two integrals that only differ in the nucleon propagator, when
applying the IBP method we can set to zero any resulting integral
in which the propagator $1/(2k\cdot N+i\epsilon)$ disappears. This
can be done because the same integral with the propagator $1/(2k\cdot N-i\epsilon)$
instead would give the exact same contribution and thus the difference
between the two is zero. Once the reduction to master integrals is
complete, we can simply substitute back the delta function in place
of the nucleon propagator.

In order to calculate the master integrals, we turned the expressions
in Appendix~\ref{app} into a representation in terms of Feynman
parameters. The procedure we then followed in most cases was to use
a Mellin-Barnes representation~\cite{Smirnov:1999gc,Tausk:1999vh} to
break up sums of Feynman parameters
raised to non-integer powers and transform the integrals into integrals
of Gamma functions over the imaginary axis. In some cases, we were
able to obtain analytical results. Otherwise, we used the \texttt{Mathematica}
packages \texttt{MB}~\cite{Czakon:2005rk} and \texttt{MBresolve}~\cite{Smirnov:2009up}
to perform a numerical calculation.

For integrals $I_{9}$, $I_{10}$, $I_{14}$, $I_{15}$, $I_{27}$,
$I_{28}$, $I_{31}$, and $I_{32}$ (cf. Appendix~\ref{app}), the
Mellin-Barnes representation was too cumbersome for a numerical evaluation.
In these cases we used the \texttt{Mathematica} package \texttt{FIESTA~1.2.1}~\cite{Smirnov:2008py}
with integrators from the \texttt{CUBA} library~\cite{Hahn:2004fe}
to perform numerical computations using sector decomposition \cite{Binoth:2000ps,Heinrich:2008si}.
Like \texttt{FIRE}, \texttt{FIESTA} can only process standard propagators
as input. This means that we had to use the momentum representation
of the integrals with the nucleon propagators instead of the delta
function. We separately calculated the integrals containing $1/(2N\cdot k+i\epsilon)$
and the ones containing $1/(-2N\cdot k+i\epsilon)$ instead, and added
the two results. We checked the method by computing with \texttt{FIESTA}
some integrals we had already found with \texttt{MB} and \texttt{MBresolve}.
The results always agreed.

There was one case, integral $I_{32}$, where the \texttt{FIESTA}
result for the integral with $1/(-2N\cdot k+i\epsilon)$ was numerically
unstable. Fortunately, in this case we could find a representation
in terms of Feynman parameters that we could compute directly using
\texttt{CUBA}, without further treatment. This was possible because
the integral is finite, and the representation was free of spurious
divergences. We cross-checked this result using a beta version
of \texttt{FIESTA~2}~\cite{Smirnov:2009pb}, which did not produce the
instabilities we encountered in the former version.

We performed an additional cross-check of our results by changing
the basis of integrals. To do this, we took one of the integrals we
computed with \texttt{FIESTA} and used the IBP method to express it
in terms of a similar integral of our choice (same as the original
one, but with some propagator(s) raised to different powers) plus
other master integrals we already knew. We then computed the new integral
with \texttt{FIESTA} and checked if the final result for the Lamb
shift (or for individual diagrams) agreed with the calculation in
the old basis. Since changing the basis modifies the coefficients
of all the integrals involved in the change, the agreement of the
results obtained with different bases is a very good cross-check of
our calculations.

This cross-check was performed for several integrals. In particular,
we changed integrals $I_{19}$ and $I_{27}$, which are the ones limiting
our precision, and integrals $I_{15}$ and $I_{32}$. Since the last
two integrals contain most of the propagators for integral types $F$
and $G$, the corresponding changes of basis affect the coefficients
of most of the other integrals of the respective type.

\section{Results}
\label{res}

Our final results for the separate contributions from the vacuum-polarization
diagrams of Fig.~\ref{vacuum} and the diagrams $a$--$s$ of Fig.~\ref{other}
are
\begin{eqnarray}
  \delta E_{vac.} & = & \frac{\al^{2}(Z\al)^{5}}{\pi
    n^{3}}\lp\frac{\mu}{m}\rp^{3}m\,[0.86281422(3)]\,,\label{vac}\\
  \delta E_{a-s} & = & \frac{\al^{2}(Z\al)^{5}}{\pi
    n^{3}}\lp\frac{\mu}{m}\rp^{3}m\,[-7.72381(4)]\,.\label{as}
\end{eqnarray} 
The best results so far for the vacuum polarization diagrams and
for diagrams $a$--$s$ have been published in \cite{Pachucki:1993zz}
(cf. \cite{Eides:2000xc} for references of partial results)
and \cite{Eides:1995ey}, respectively. Our results are compatible
with them and improve the precision by two orders of magnitude in
the case of $\delta E_{vac.}$ and a little over one order of magnitude
for $\delta E_{a-s}$.

The total result reads
\begin{equation}
  \delta E=\frac{\al^{2}(Z\al)^{5}}{\pi
    n^{3}}\lp\frac{\mu}{m}\rp^{3}m\,[-6.86100(4)]\,,\label{total}
\end{equation} 
and the corresponding energy shifts for the $1S$ and the $2S$ states in
hydrogen are
\begin{eqnarray}
  \delta E_{1S} & = & -296.866(2)\,\text{kHz}\,,\\
  \delta E_{2S} & = & -37.1082(3)\,\text{kHz}\,.
\end{eqnarray}

\begin{table}[tb]
  \caption{\label{tvac} Comparison between our results for the different
    vacuum-polarization sets (in Fried-Yennie gauge) and those of
    \cite{Pachucki:1993zz,Eides:1996uj}. Numbers ending in an ellipsis
    indicate an analytic result, which we show in Appendix \ref{appB}.}
  \begin{ruledtabular}
    \begin{tabular}{lll}
      Set  & This paper  & Refs. \cite{Pachucki:1993zz,Eides:1996uj} \\
      \hline
      I & $-0.07290996446926(4)$ & $-0.0729098(3)$\\
      II & 0.61133839226\dots  & $0.61133839226\dots$\\
      III & 0.50814858506\dots & $0.50814858506\dots$\\
      IV & $-0.12291623(3)$ & $-0.122915(3)$\\
      V & $-23/378$ & $-23/378$\\
    \end{tabular}
  \end{ruledtabular} 
\end{table}

\begin{table}[tb]
  \caption{\label{tother} Comparison between our results for diagrams
    $a$--$s$ (in Fried-Yennie gauge) and those of
    \cite{Eides:1995ey}. Numbers ending in an ellipsis indicate an
    analytic result, which we show in Appendix \ref{appB}.}
  \begin{ruledtabular}
    \begin{tabular}{lll}
      Diagram  & This paper  & Ref. \cite{Eides:1995ey}\\
      \hline
      $a$ & 0 & 0\\
      $b$ & 2.955090809\dots & 2.9551(1)\\
      $c$ & $-2.22312657\dots$ & $-2.2231(1)$\\
      $d$ & $-5.2381153272259(2)$ & $-5.238023(56)$\\
      $e$ & 5.0561650638185(4) & 5.056278(81)\\
      $f$ & $6\ln2-207/40$ & $-1.016145(21)$\\
      $g$ & $6\ln2-147/80-\pi^{2}/4$ & $-0.1460233(52)$\\
      $h$ & 153/80 & 153/80\\
      $i$ & $-5.51731(2)$ & $-5.51683(34)$\\
      $j$ & $-7.76838(1)$ & $-7.76815(17)$\\
      $k$ & 1.9597582447795(2) & 1.959589(33)\\
      $l$ & 1.74834(4) & 1.74815(38)\\
      $m$ & 1.87510512(6) & 1.87540(17)\\
      $n$ & $-1.30570289(7)$ & $-1.30584(18)$\\
      $o$ & $-12.06904(9)$ & $-12.06751(47)$\\
      $p$ & 6.13815(1) & 6.13776(25)\\
      $q$ & $-7.52425(2)$ & $-7.52453(34)$\\
      $r$ & 14.36962(7) & 14.36733(44)\\
      $s$ & $-0.9304766935602(5)$ & $-0.930268(72)$\\
    \end{tabular}
  \end{ruledtabular} 
\end{table}

Choosing the Fried-Yennie gauge \cite{Fried:1958zz,Adkins:1993qm},
we also compared the results from the different sets of vacuum polarization
diagrams with those of \cite{Pachucki:1993zz,Eides:1996uj}, and the
results from the individual diagrams $a$--$s$ with those of \cite{Eides:1995ey}.
Our results for the vacuum polarization graphs and diagrams $a$--$s$
are presented in Tables \ref{tvac} and \ref{tother}, respectively.
All numbers in the tables are to be multiplied by the prefactor
$\al^{2}(Z\al)^{5}/(\pi n^{3})(\mu/m)^{3}m$
(note the difference in normalization in \cite{Pachucki:1993zz}).

We found new analytic results for four diagrams. The results for diagrams
$f$ and $g$ are given in Table~\ref{tother}, while the results
for diagrams $b$ and $c$, being too lengthy for the table, are presented
in Appendix~\ref{appB}. For completeness, the known analytic results
for sets II and III of the vacuum polarization diagrams are given
in Appendix~\ref{appB} as well.

It should be mentioned that the errors of the results in Eqs. (\ref{vac}),
(\ref{as}), and (\ref{total}) are not obtained from the sum of the
errors of the diagrams in Tables \ref{tvac} and \ref{tother}. Once
we decompose the problem into the calculation of master integrals,
the diagrams are no longer independent, as the same master integral
contributes to several different diagrams. Thus, to find the error
of our total result, we first sum all diagrams and then sum all the
errors of the integrals in quadrature.

We found discrepancies between our results for diagrams $a$--$s$
and those of \cite{Eides:1995ey}. Most of the central values in the
second and third column of table \ref{tother} lie between $1\sigma$
and $2\sigma$ away from each other, but in the case of diagrams $o$
and $s$ the difference is around $3\sigma$, and for diagrams $k$
and $r$, it reaches $5\sigma$ (we take as $\sigma$ the errors of
individual diagrams in the third column). We should stress again that
our calculation is done using dimensional regularization while the
study of Ref.~\cite{Eides:1995ey} was performed in four dimensions.
Even though all the individual diagrams are finite, one can imagine
situations where the two regularization methods give different partial
results. However, we do not observe significant cancellations in the
sum of the differences. Thus, it seems the differences are real although
practically negligible; their sum is very small and amounts to $10^{-3}$,
which is the error estimate in \cite{Eides:1995ey}. Thus our results
agree within that error.

\section{Summary}

We have applied particle theory methods to compute, in dimensional
regularization and a general covariant gauge, the corrections of order
$\al^{2}(Z\al)^{5}$ to the Lamb shift. We have made use of IBP techniques
to reduce the problem of computing all the necessary Feynman diagrams
to the simpler problem of computing 32 scalar integrals. Mellin-Barnes
integral representations and sector decomposition have then allowed
us to obtain analytic results for some of these integrals, and good
numerical results for the rest. With this, we have been able to reproduce
and improve the results from previous calculations. The techniques
used here are quite general and can be applied to other multi-loop
problems in atomic physics.

\begin{acknowledgments}
We are grateful to A.V. and V.A.~Smirnov for providing us with a new
version of \texttt{FIESTA} prior to publication. We thank M.I.~Eides for
useful comments. This work was supported by the Natural Sciences and
Engineering Research Council of Canada. The work of J.H.P. was supported
by the Alberta Ingenuity Foundation. The Feynman diagrams were drawn
using \texttt{Axodraw}~\cite{Vermaseren:1994je} and
\texttt{Jaxodraw 2}~\cite{Binosi:2008ig}.
\end{acknowledgments}

\appendix

\begin{widetext} 

\section{Results for the master integrals}
\label{app}

In section \ref{eval} we presented our method of calculating the
corrections to the Lamb shift, which differs significantly from the
methods used in previous calculations. One important difference is
the reduction of diagrams to master integrals. Here we present our
results for all master integrals.

\begin{figure}[t]
  \includegraphics[width=\columnwidth]{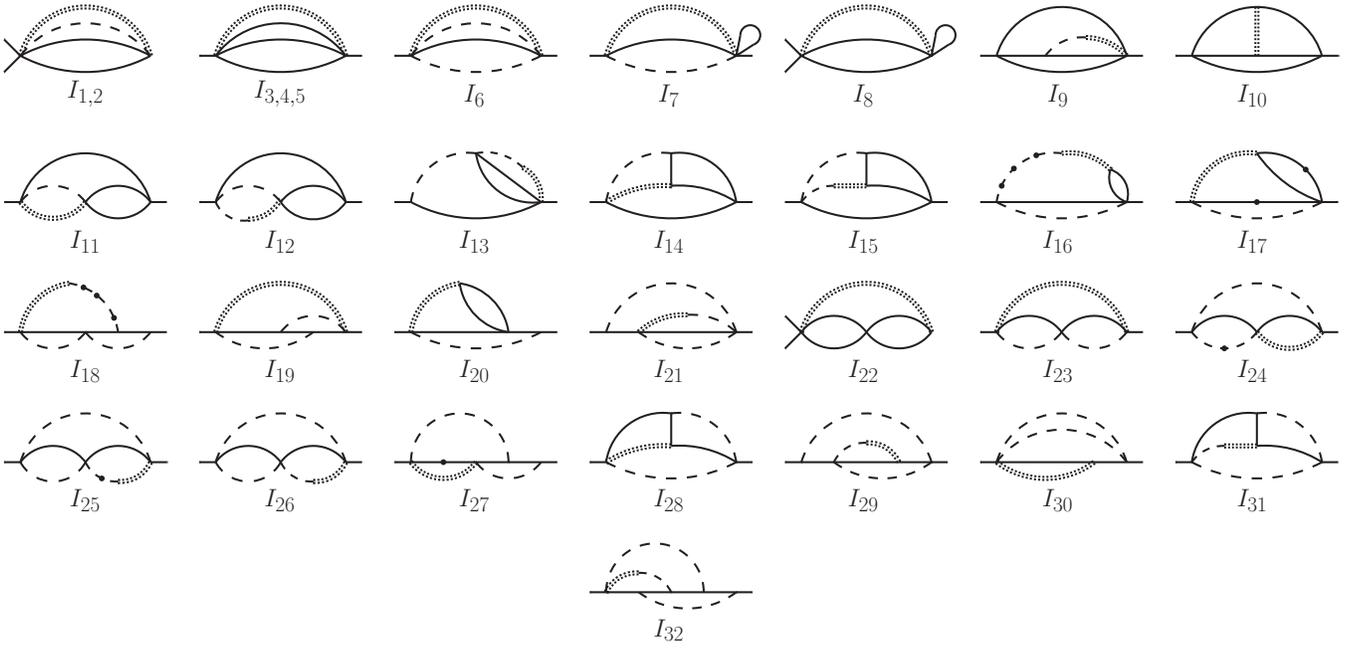}
  \caption{\label{masters} A graphic representation of the 32 master
    integrals. Solid and dashed lines represent massive and massless
    scalar propagators, respectively. The dotted double line denotes the
    delta function. A dot on a line signifies that the propagator is
    raised to a higher power. The external lines indicate the momentum
    $p$ of the electron flowing in and out of the diagram. The first two
    diagrams represent different integrals that differ only by a term in
    the numerator.}
\end{figure}

The set of master integrals is represented in Fig. \ref{masters}.
We have two different types of integrals. In Euclidean space, they
are defined as:
\begin{eqnarray}
  &  & F(\nu_{1},\nu_{2},\nu_{3},\nu_{4},\nu_{5},\nu_{6},\nu_{7},\nu_{8})=\frac{e^{3\gamma_{E}\epsilon}}{\lp\pi^{D/2}\rp^{3}}\int\frac{d^{D}k_{1}\, d^{D}k_{2}\, d^{D}k_{3}\;2\pi\,\delta(k_{2}^{0})}{(k_{1}^{2})^{\nu_{1}}\,(k_{2}^{2})^{\nu_{2}}\,(k_{3}^{2})^{\nu_{3}}\,[(k_{1}+p)^{2}+1]^{\nu_{4}}}\nn\\
 &  & \times\frac{1}{[(k_{1}+k_{2}+p)^{2}+1]^{\nu_{5}}\,[(k_{1}+k_{2}+k_{3}+p)^{2}+1]^{\nu_{6}}\,[(k_{2}+k_{3}+p)^{2}+1]^{\nu_{7}}\,[(k_{3}+p)^{2}+1]^{\nu_{8}}}\,,\\
 &  & G(\nu_{1},\nu_{2},\nu_{3},\nu_{4},\nu_{5},\nu_{6},\nu_{7})=\frac{e^{3\gamma_{E}\epsilon}}{\lp\pi^{D/2}\rp^{3}}\int\frac{d^{D}k_{1}\, d^{D}k_{2}\, d^{D}k_{3}\;2\pi\,\delta(k_{1}^{0})}{(k_{1}^{2})^{\nu_{1}}\,(k_{2}^{2})^{\nu_{2}}\,(k_{3}^{2})^{\nu_{3}}\,[(k_{1}+k_{2}+p)^{2}+1]^{\nu_{4}}}\nn\\
 &  & \times\frac{1}{[(k_{1}+k_{2}+k_{3}+p)^{2}+1]^{\nu_{5}}\,[(k_{2}+k_{3}+p)^{2}+1]^{\nu_{6}}\,[(k_{3}+p)^{2}+1]^{\nu_{7}}}\,,
\end{eqnarray}
where $D=4-2\epsilon$, and $p=(i,\vec{0})$ is the momentum of the
electron. The mass of the electron has been set equal to one for convenience
and can be easily restored from the dimension of the integral. The
factor $e^{3\gamma_{E}\epsilon}$, where $\gamma_{E}$ is the Euler-Mascheroni
constant, has been introduced to suppress the dependence of the results
on this constant.

With these definitions, our results for the master integrals are:
\begin{eqnarray}
I_{1} & = & F(1,0,0,0,0,1,0,1)=2e^{3\gamma_{E}\epsilon}\,\frac{\Gamma(1-\epsilon)\Gamma^{2}\left(-\frac{3}{2}+2\epsilon\right)\Gamma\left(-\frac{1}{2}+\epsilon\right)\Gamma\left(-\frac{5}{2}+3\epsilon\right)}{\Gamma(-3+4\epsilon)}\,,\label{I1}\\
I_{2} & = & F(1,0,-1,0,0,1,0,1)=-2I_{1}\,,\label{I2}\\
I_{3} & = & F(0,0,0,0,1,1,1,0)=-96.174642407742494299(1)-3003.97283051743374945(1)\epsilon\nn\\
 &  & -16370.644886761701890(1)\epsilon^{2}-204040.09217878970569(1)\epsilon^{3}+\mathcal{O}(\epsilon^{4})\,,\label{I3}\\
I_{4} & = & F(-1,0,0,1,0,1,0,1)=128.23285654365665907(1)+4005.2971073565783326(1)\epsilon\nn\\
 &  & +21827.526515682269186(1)\epsilon^{2}+272053.45623838627426(1)\epsilon^{3}+\mathcal{O}(\epsilon^{4})\,,\label{I4}\\
I_{5} & = & F(0,-1,0,0,1,1,1,0)=213.37528929773859515(1)-1789.0076495990746772(1)\epsilon+\mathcal{O}(\epsilon^{2})\,,\label{I5}\\
I_{6} & = & F(1,0,1,0,0,1,0,0)=2\sqrt{\pi}e^{3\gamma_{E}\epsilon}\,\frac{\Gamma^{2}(1-\epsilon)\Gamma\left(-\frac{5}{2}+3\epsilon\right)\Gamma\left(-\frac{3}{2}+2\epsilon\right)\Gamma\left(\frac{9}{2}-5\epsilon\right)}{\Gamma(2-2\epsilon)\Gamma(3-3\epsilon)}\,,\\
I_{7} & = & F(1,0,0,0,1,0,0,1)=2\sqrt{\pi}e^{3\gamma_{E}\epsilon}\,\frac{\Gamma(-1+\epsilon)\Gamma\left(-\frac{3}{2}+2\epsilon\right)\Gamma\left(\frac{5}{2}-3\epsilon\right)\Gamma(-\frac{1}{2}+\epsilon)}{\Gamma(2-2\epsilon)}\,,\\
I_{8} & = & F(0,0,0,0,1,0,1,1)=2\sqrt{\pi}e^{3\gamma_{E}\epsilon}\,\frac{\Gamma(-1+\epsilon)\Gamma\left(-\frac{3}{2}+2\epsilon\right)\Gamma^{2}\left(-\frac{1}{2}+\epsilon\right)}{\Gamma(-1+2\epsilon)}\,,\\
I_{9} & = & F(0,1,0,0,1,1,1,1)=-\frac{8\pi^{2}}{\epsilon}-257.35053226188(1)-2952.9668342496406(4)\epsilon+\mathcal{O}(\epsilon^{2})\,,\\
I_{10} & = & F(0,0,0,1,1,1,1,1)=-420.49901(1)+1860.837(4)\epsilon+\mathcal{O}(\epsilon^{2})\,,\\
I_{11} & = & F(1,0,0,0,1,1,0,1)=2^{3-4\epsilon}\pi e^{3\gamma_{E}\epsilon}\,\frac{\Gamma\lp\epsilon-\frac{1}{2}\rp}{\cos\lp2\pi\epsilon\rp}\left[2\frac{\Gamma\lp\frac{5}{2}-3\epsilon\rp\Gamma\lp\epsilon\rp}{\Gamma\lp4-4\epsilon\rp}\,{}_{3}F_{2}\lp1,\frac{5}{2}-3\epsilon,\epsilon;\frac{3}{2},4-4\epsilon;1\rp\right.\nonumber \\
 &  & \left.-\sqrt{\pi}\frac{\Gamma\lp1-\epsilon\rp\Gamma\lp3\epsilon-\frac{3}{2}\rp}{\Gamma\lp\frac{5}{2}-2\epsilon\rp\Gamma\lp2\epsilon\rp}\,{}_{3}F_{2}\lp1,1-\epsilon,3\epsilon-\frac{3}{2};\frac{5}{2}-2\epsilon,2\epsilon;1\rp\right]\,,\label{I11}\\
I_{12} & = & F(1,1,0,0,1,1,0,1)=-\frac{4\pi^{2}}{\epsilon}-24\pi^{2}-\frac{4\pi^{4}}{3}+\pi^{2}\lp-116-\frac{59}{3}\pi^{2}+32\ln2+100\zeta(3)\rp\epsilon+\mathcal{O}(\epsilon^{2})\,,\\
I_{13} & = & F(1,1,0,1,0,1,0,1)=-263.74028719521945979(1)+1741.1125810306205720(1)\epsilon+\mathcal{O}(\epsilon^{2})\,,\\
I_{14} & = & F(1,0,0,0,1,1,1,1)=-362.8560(1)+\mathcal{O}(\epsilon)\,,\\
I_{15} & = & F(1,1,0,0,1,1,1,1)=36.969282(2)+\mathcal{O}(\epsilon)\,,\\
I_{16} & = & F(1,4,0,0,1,0,1,1)=\pi^{2}\left(-\frac{513}{128\epsilon}-\frac{11077}{768}-\frac{571}{16}\ln2+32\sqrt{5}\ln\left(\frac{1+\sqrt{5}}{2}\right)\right)\nn\\
 &  & -1889.3810189605726842(1)\epsilon-2199.2559561980712031(1)\epsilon^{2}+\mathcal{O}(\epsilon^{3})\,,\\
I_{17} & = & F(1,0,0,0,2,0,2,1)=\frac{32\pi^{2}}{\sqrt{5}}\ln\left(\frac{1+\sqrt{5}}{2}\right)-683.43054120051764110(1)\epsilon\nn\\
 &  & +5647.2496334930969112(1)\epsilon^{2}+\mathcal{O}(\epsilon^{3})\,,\\
I_{18} & = & F(1,4,1,0,1,0,1,1)=\pi^{2}\left[\frac{343}{512\epsilon}+\frac{125257}{46080}-\frac{2}{3}\pi^{2}+\frac{169}{192}\ln2+16\ln^{2}2-48\ln^{2}\lp\frac{1+\sqrt{5}}{2}\rp\right]+\mathcal{O}(\epsilon)\,,\\
I_{19} & = & F(1,0,1,0,1,1,1,0)=-293.4480(2)+\mathcal{O}(\epsilon)\,,\\
I_{20} & = & F(1,0,0,1,1,0,1,1)=\pi^{2}\left[-\frac{2}{\epsilon}-2-\frac{8}{3}\pi^{2}+16\sqrt{5}\ln\lp\frac{1+\sqrt{5}}{2}\rp+32\ln^{2}\lp\frac{1+\sqrt{5}}{2}\rp\right]\nn\\
 &  & +1394.0754186124348755(1)\epsilon+\mathcal{O}(\epsilon^{2})\,,\label{I20}\\
I_{21} & = & F(1,1,1,0,0,1,0,1)=2\pi^{2}-\frac{4\pi^{4}}{3}+\pi^{2}\lp44-4\pi^{2}+80\zeta(3)\rp\epsilon+\mathcal{O}(\epsilon^{2})\,,\\
I_{22} & = & F(0,0,0,1,1,0,1,1)=-\frac{128\pi^{2}}{3}-\pi^{2}\left(\frac{1792}{3}+\frac{256}{3}\pi-\frac{1024}{3}\ln2\right)\epsilon+\mathcal{O}(\epsilon^{2})\,,\label{I22}\\
I_{23} & = & F(1,0,1,0,1,0,1,0)\nonumber \\
 & = & \frac{2\pi^{3}e^{3\gamma_{E}\epsilon}\,}{\Gamma^{2}\lp2-2\epsilon\rp\Gamma\lp\frac{3}{2}-\epsilon\rp}\left[2^{2\epsilon-2}\frac{\Gamma\lp1-\epsilon\rp\Gamma\lp\epsilon-\frac{1}{2}\rp}{\sin\lp\pi\epsilon\rp\cos\lp4\pi\epsilon\rp}\lp\frac{\Gamma\lp\frac{7}{2}-5\epsilon\rp\sin\lp\pi\epsilon\rp}{\Gamma\lp\frac{5}{2}-3\epsilon\rp\sin\lp2\pi\epsilon\rp\cos\lp3\pi\epsilon\rp}-\frac{\Gamma\lp3\epsilon-\frac{3}{2}\rp}{\Gamma\lp5\epsilon-\frac{5}{2}\rp\cos\lp2\pi\epsilon\rp}\rp\right.\nonumber \\
 &  & +\frac{\sqrt{\pi}\,\Gamma\lp2-2\epsilon\rp\Gamma\lp2\epsilon-\frac{1}{2}\rp}{\Gamma\lp2-\epsilon\rp\Gamma\lp3\epsilon-\frac{1}{2}\rp\sin\lp\pi\epsilon\rp\cos\lp\pi\epsilon\rp\cos\lp3\pi\epsilon\rp}\,{}_{3}F_{2}\lp1,2-2\epsilon,2\epsilon-\frac{1}{2};2-\epsilon,3\epsilon-\frac{1}{2};1\rp\nonumber \\
 &  & \left.-\frac{2^{1-2\epsilon}\pi\,\Gamma\lp\frac{5}{2}-3\epsilon\rp}{\Gamma\lp\frac{5}{2}-2\epsilon\rp\Gamma\lp\frac{1}{2}+\epsilon\rp\sin\lp2\pi\epsilon\rp\cos\lp\pi\epsilon\rp\cos\lp2\pi\epsilon\rp}\,{}_{3}F_{2}\lp1,\frac{5}{2}-3\epsilon,\epsilon;\frac{5}{2}-2\epsilon,2\epsilon;1\rp\right]\,,\\
I_{24} & = & G(0,1,2,1,0,1,0)=\frac{2\pi^{2}}{\epsilon}-162.745878930257(1)+640.681562239(2)\epsilon\nn\\
 &  & -9490.745115169417(3)\epsilon^{2}+\mathcal{O}(\epsilon^{3})\,,\\
I_{25} & = & G(2,1,1,1,0,1,0)=-\frac{4\pi^{2}}{\epsilon}-192.3546921335253(1)-2297.18352848038(1)\epsilon\nn\\
 &  & -10356.58582995624(1)\epsilon^{2}+\mathcal{O}(\epsilon^{3})\,,\\
I_{26} & = & G(1,1,1,1,0,1,0)=-\frac{4\pi^{2}}{\epsilon}-244.4995291143211(3)-2339.54007847666(2)\epsilon+\mathcal{O}(\epsilon^{2})\,,\\
I_{27} & = & G(0,1,1,2,0,1,1)=136.8086023(2)-907.048(2)\epsilon+\mathcal{O}(\epsilon^{2})\,,\\
I_{28} & = & G(0,1,1,1,1,1,0)=-280.62418(1)+734.494(1)\epsilon+\mathcal{O}(\epsilon^{2})\,,\\
I_{29} & = & G(1,1,1,0,1,1,1)=118.63826101784(1)+\mathcal{O}(\epsilon)\,,\\
I_{30} & = & G(0,1,1,0,1,1,0)=-10\sqrt{\pi}e^{3\gamma_{E}\epsilon}\,\frac{\Gamma\left(\epsilon\right)\Gamma^{2}\lp1-\epsilon\rp\Gamma\left(\frac{5}{2}-5\epsilon\right)\Gamma\lp-\frac{3}{2}+3\epsilon\rp}{\Gamma\lp2-2\epsilon\rp\Gamma\left(\frac{7}{2}-4\epsilon\right)}\nn\\
 &  & \times{}_{3}F_{2}\left(\frac{7}{2}-5\epsilon,\frac{3}{2}-\epsilon,-\frac{1}{2}+\epsilon;\frac{7}{2}-4\epsilon,\frac{1}{2}+\epsilon;1\right)\,,\\
I_{31} & = & G(1,1,1,1,1,1,0)=49.3616(1)+\mathcal{O}(\epsilon)\,,\\
I_{32} & = & G(1,1,1,1,1,1,1)=26.272804(6)+291.1097(1)\epsilon+\mathcal{O}(\epsilon^{2})\,,
\end{eqnarray}
\end{widetext}
where $\zeta$ denotes Riemann's zeta function, and
${}_{3}F_{2}$ is a generalized hypergeometric function. The latter
can be expanded in $\epsilon$ with the help of the \texttt{Mathematica}
package \texttt{HypExp~2}~\cite{Huber:2007dx}.

The relation between integrals $I_{1}$ and $I_{2}$ expressed in
Eq.~(\ref{I2}) is not evident when looking at their respective representations.
This relation becomes clear when checking the cancellation of the
gauge-parameter dependence in the sum of all diagrams. If the 32 integrals
presented here were an irreducible basis, the gauge dependence of
the coefficient of each integral should vanish independently. However,
this does not happen with the coefficients of integrals $I_{1}$ and
$I_{2}$, which means that the integrals are connected. Demanding
the cancellation of the gauge dependence yields Eq.~(\ref{I2}).
We checked this relation by computing explicitly the analytic solution
for $I_{2}$.

There appears to be also a relation between integrals $I_{3}$ and
$I_{4}$, although the gauge dependence does not give us any hint
in this case. By demanding the cancellation of poles in several
diagrams, one can find the following relation between the first three
terms of $I_{3}$ and $I_{4}$,
\begin{equation}
  I_{4}=-\frac{4}{3}I_{3} + \mathcal{O}(\epsilon^3)\,.
\end{equation}
The relation, however, seems to be valid to all orders in the $\epsilon$
expansion. We checked it numerically up to order $\epsilon^{4}$,
but we could not find an analytic proof for it.

Integrals $I_{3}$--$I_{5}$, $I_{12}$, $I_{13}$, $I_{16}$, $I_{17}$,
and $I_{20}$--$I_{22}$ can be represented as a one-fold Mellin-Barnes
integral. We only show numerical results with 20-digit precision,
which is more than enough for our purposes. However, these integrals
can be easily evaluated with a precision of 100 digits or more. With
this kind of precision it is possible to find analytical results,
using the \texttt{PSLQ} algorithm~\cite{pslq}. In this way, we determined
the $\mathcal{O}(\epsilon)$ term of integrals $I_{12}$ and $I_{21}$.

The analytic expression for the $\mathcal{O}(\epsilon^{0})$ term
in $I_{20}$ was obtained using the analytic result for set II of
vacuum-polarization diagrams presented in \cite{Eides:1996uj}. Likewise,
the analytic expression for the $\mathcal{O}(\epsilon)$ term in $I_{22}$
was extracted from the analytic result for set III found in \cite{Pachucki:1993zz}.
As mentioned above, we were able to numerically calculate these integrals
to 100-digit precision and confirm the analytic expressions with \texttt{PSLQ}.





\section{Analytic results}
\label{appB}

Here we show the analytic results for diagrams $b$ and $c$ from
Fig. \ref{other}:
\begin{eqnarray}
  \lefteqn{\text{Diagram }b =} && \nn\\
  && \frac{111}{8} - \pi^{2}-9\ln2+24\ln^{2}2 +
  \frac{48}{\sqrt{5}}\ln\lp\frac{1+\sqrt{5}}{2}\rp \nn\\
  && -72\ln^{2}\lp\frac{1+\sqrt{5}}{2}\rp,\\
  \lefteqn{\text{Diagram }c =} && \nn\\
  && -\frac{352897}{27000} + \frac{31}{45}\pi^{2} - \frac{643}{225}\ln2
  - \frac{248}{15}\ln^{2}2 \nn\\
  && +\frac{104}{9\sqrt{5}}\ln\lp\frac{1+\sqrt{5}}{2}\rp +
  \frac{248}{5}\ln^{2}\lp\frac{1+\sqrt{5}}{2}\rp.
\end{eqnarray}

For completeness, we also give here the analytic results for sets
II and III of the vacuum polarization diagrams, found in \cite{Eides:1996uj}
and \cite{Pachucki:1993zz}, respectively:
\begin{eqnarray}
  \text{Set II} & = &
  \frac{67282}{6615} - \frac{2}{9}\pi^{2} + \frac{628}{63}\ln2 \nn\\
  && -\frac{872}{63}\sqrt{5}\ln\lp\frac{1+\sqrt{5}}{2}\rp +
  \frac{8}{3}\ln^{2}\lp\frac{1+\sqrt{5}}{2}\rp\,, \nn\\
  \\
  \text{Set III} & = & \frac{15647}{13230}-\frac{25}{63}\pi+\frac{52}{63}\ln2\,.
\end{eqnarray}

\end{document}